# A Systematic Comparison of In-Context Learning Strategies for Peptide Property Prediction in Low-Data Regimes

**Jiantao Zhao[1*], Andrei Voinea[1*], Mathijs F.J. Mabesoone[1†]**

[1]Institute for Molecules and Materials, Radboud University, Heyendaalseweg 135, 6525AJ Nijmegen, The Netherlands

*These authors contributed equally †email: mathijs.mabesoone@ru.nl

## Abstract

Peptide sequence-property relationships are often highly non-linear, but can be learned with machine learning models and exploited for the design of new materials and bioactive compounds. Various methods based on transformer architectures have become available for molecular property predictions. These foundation models promise straightforward inference and prediction of properties of new materials through in-context learning. Benchmarks of individual model architectures exist, but a systematic comparison of various foundation models against classical baselines on similar datasets is lacking. Here, we provide a systematic evaluation of the predictive performance of various in-context learning strategies. We compare several large language models, tabular foundation models, and typed decision models against an XGBoost baseline. We evaluate predictive performance on binary and multiclass classification tasks on pH-dependent surface activity, IR spectral signatures, self-assembly, and anti-cancer activities. Our results show that ICL is competitive, yet does not significantly outperform classical baselines, and performance varies between models and tasks. Our findings further suggest that language models use both in context examples and knowledge acquired during pretraining to infer properties of unseen samples. This finding indicates that pretrained models can leverage the existing literature for predictions, but may also bias their predictions towards established sequence-property relationships.

## Introduction

The properties of peptides are distinctly dependent on their sequence. Myriad non-linear interactions between residues drive the emergence of secondary structure elements and can cause counterintuitive effects also in short peptides.[1–3] These non-linear interactions make straightforward, linear interpretation of sequence-property relationships often challenging, and create a necessity to use more advanced statistical methods to understand these relationships and predict properties of new peptides.[4–8] Such non-linear mapping of sequence to property can ideally be done with machine learning (ML) models.

The choice for a specific machine learning model architecture typically depends on specific details of the data on which the model will be trained. Many areas of peptide science are constrained by synthetic throughput, and typically are in a low data regime, with $10^1$–$10^3$ labeled peptide sequences available.[9] Whereas models to predict properties that can readily be measured with high-throughput screens, such as antimicrobial or other biological activities, are often trained on thousands of datapoints,[10,11] datasets on many other properties, such as surface activity,[6] IR spectral features,[7] aggregation,[4,8] and peptide catalysis[12] are considerably smaller in size. In these cases, deep neural networks, which often achieve superior performance in high-data regimes, are less suited.[13] Other supervised ML techniques are needed to learn sequence-property relationships and predict properties of uncharacterized peptides.

Most common classical supervised machine learning models rely on various inductive biases in model design and data representation. For instance, the choice for molecular embedding, kernel functions, and model hyperparameters can all drastically affect model performance, and typically require task-specific optimization. Recently, large pretrained foundation models have emerged as a promising alternative for molecular property prediction.[14,15] Rather than relying on specific inductive biases, these models leverage their large-scale pretraining to identify correlations from the data and enable prediction of unseen samples. These models keep their weights fixed, so there are no training hyperparameters to tune, and effort shifts from model engineering to the data.

Whereas earlier examples focused on fine-tuning large language models on hand-curated datasets,[14,16,17] modern frontier models often give good performance without fine-tuning, and can infer correlations with some, or even no examples provided in the language model prompt.[18] Trends and correlations from the large-scale pretraining enable this few-shot or zero-shot prediction through what is known as in-context learning.[19–21] The in-context learning that large language models enable provides a straightforward and hassle-free approach to predict properties of new molecules and materials,[21,22] but running large language models comes at considerable computational cost compared to classical ML models. Admittedly, this comparatively large computational costs of LLMs make them unattractive for widespread deployment.

The large computational overhead that burdens large language models is mitigated by more specialized foundation models that are specifically designed for numerical and Boolean prediction. Tabular foundation models (TFMs) are gaining increased interest for such tasks.[23,24] These transformer-based models are trained on a large amount of tabular data. Large-scale pretraining on real-world and synthetic data enables these models to relate patterns in the in-context examples to previously learned associations, allowing them to infer properties of unseen samples. Analogous to large language models, TFMs require little to no hyperparameter optimization or inductive biases, are robust to shifts in the input distribution, and already function well in the low data regime.

Very recently, typed decision models (TDMs),[25] such as Jev,[26] have emerged as interesting alternative approaches to combine pretraining on large bodies of natural language, including scientific texts, with fast and inexpensive inference. The output of typed decision models is not generated through autoregressive token generation, but is simply a probability distribution of predefined class labels. The combination of fast and inexpensive inference, a lack of required hyperparameter finetuning, and freeform data representation make TDMs interesting as potential platforms for the prediction of molecular properties.

Hence, the increasingly lower entry barriers for using pretrained foundation models appear to make in context learning an attractive and simple strategy to predict properties of molecules and materials. However, the distinct design principles underlying these models raise various questions on the capacity of these models to predict new material properties for diverse molecular classes, including peptides. While various studies have reported benchmarks and comparisons of specific foundation models against classical baselines,[22,27–29] a systematic comparison of various ICL approaches on the same datasets is required to reliably assess relative performance of these new approaches to machine learning for chemistry and materials.

Here, we systematically investigate the use of in-context learning for peptide property prediction in low data regimes. We compare general-purpose LLMs, TFMs, and TDMs with conventional supervised machine learning approaches across a diverse collection of peptide classification tasks covering self-assembly, surface activity, spectroscopy-derived structural descriptors, and biological activity. We investigate classification on a range of different properties and perform a series of control experiments to assess the limitations of ICL for peptide property prediction. Through this benchmarking study, we aim to establish whether and how in-context learning can serve as a practical alternative to classical ML methods for peptide materials research.

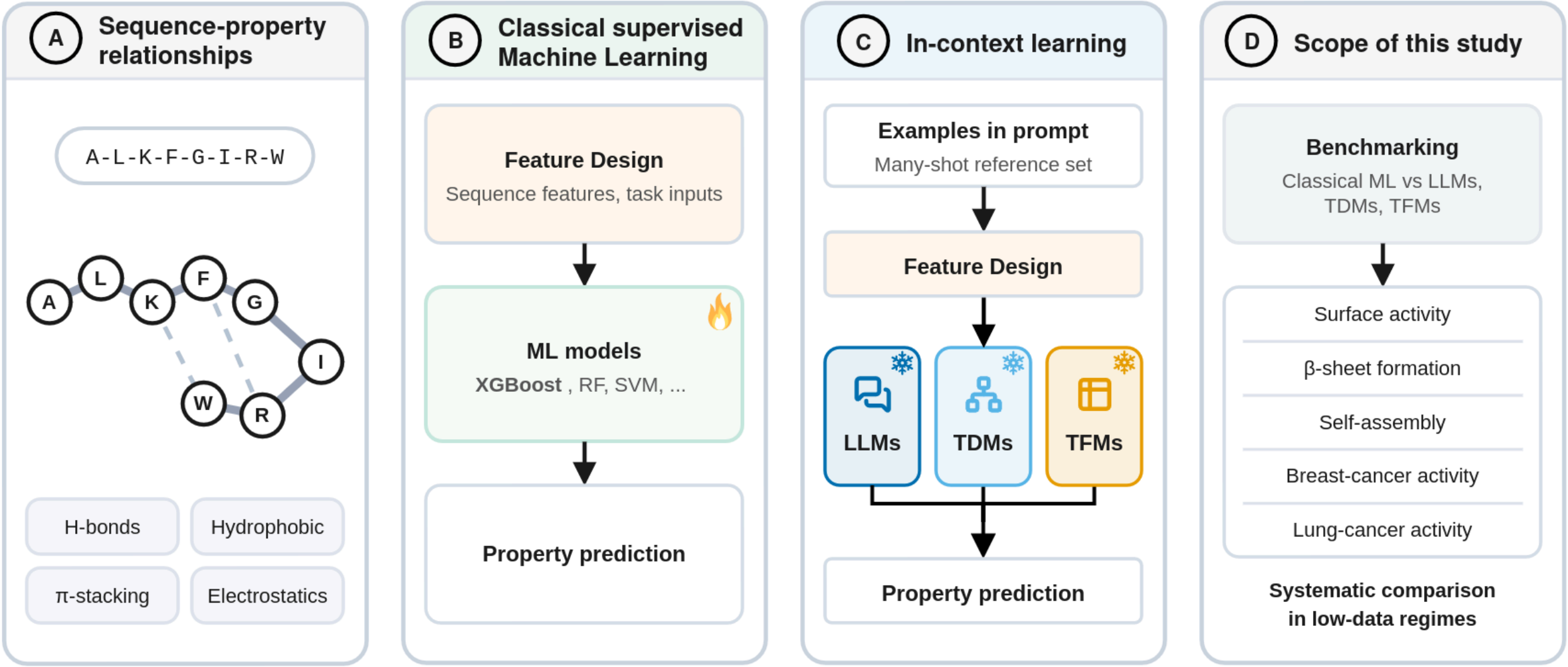

**Figure 1.** Classical machine learning and in-context learning for peptide property prediction. (a) Non-linear sequence–property relationships depend on interactions such as hydrogen bonds, hydrophobic contacts, $\pi$-stacking, and electrostatics. (b) The classical route uses a task-specific encoding and a fitted ML model. (c) In-context learning gives labelled reference examples from the full or class-balanced training set (many-shot) and a target sample to a frozen LLM, TDM, or TFM. (d) Scope: eight models, five tasks, and a comparison of classical ML against LLMs, TDMs, and TFMs in low-data regimes. Controls and modes cover the endpoint models only.

# Methods

We established predictive performance for various in-context learning models on various datasets. To determine whether a model uses the supplied context or its pretraining prior, we give every model the same five tasks, frozen splits, and reference sets. Only the model, its input representation, and the prompt change. We compare three LLMs, two typed decision models, two tabular foundation models, and an XGBoost baseline. In the following sections, we describe the selected low-data regime tasks, the used models, the input representations, and the evaluation.

## 2.1 Peptide datasets

We selected five peptide classification tasks for evaluating ICL in the low-data regime. The tasks cover both physicochemical peptide properties and biological activity, and include binary and multiclass classification problems. They also differ in the information available to the model: four tasks use peptide sequence as the sole input, whereas the surface activity task additionally includes pH. This diversity allows us to assess whether differences between model classes are consistent across peptide properties, label structures, and input representations. Table 1 summarizes the datasets and the resulting sample counts used in the study.

**Table 1.** Peptide classification tasks used in this study.

| Task | Surface activity[6] | Beta-sheet formation[7] | Self-assembly[8] | Breast-cancer activity[30] | Lung-cancer activity[30] |
|---|---|---|---|---|---|
| Number of samples | 409 | 268 | 368 | 199 | 151 |
| Classes | 2 | 2 | 2 | 3 | 3 |
| Inputs | Sequence, pH | Sequence | Sequence | Sequence | Sequence |
| Full train samples | 328 | 214 | 294 | 159 | 120 |
| Test samples | 81 | 54 | 74 | 40 | 31 |
| Class-balanced training samples | 242 | 146 | 190 | 45 | 57 |

For the benchmark on the surface activity dataset, we selected the minimum reported surface tension for each peptide amphiphile headgroup sequence and pH, and followed reported surface activity: samples with a minimum surface tension below 50 mN/m were classified as active. For $\beta$-sheet formation and self-assembly, we used the experimentally assigned binary labels provided in the source datasets. The two cancer datasets were treated as separate three-class tasks, using the three original labels: experimentally inactive, moderately active, and very active. All tasks used the peptide sequence as an input, and the surface activity task additionally used pH and tail length.

We fixed the train–test partitions before model evaluation. Four datasets used a stratified 80:20 split. For surface activity, we used a stratified group split by peptide sequence to prevent observations for the same sequence at different pHs from appearing in both partitions. We evaluated both the full training sets and balanced variants obtained by random undersampling to the size of the smallest training class. These test sets retained their original class distributions.

Section S1.1 and Table S9 provide the data preparation procedures, label mappings, sample counts, split implementation, and class-balancing rules.

## 2.2 Models and input representations

We compared eight models across four model families: three LLMs at various sizes, two typed decision models, two tabular foundation models, and an XGBoost baseline. Table 2 summarizes the models and their families.

**Table 2.** Models compared in this study.

| Model | Type | Model size | Cost per 1M in/out tokens |
|---|---|---|---|
| GPT-6.1 Sol | LLM | N/A | 2.00$ / 10.00$ |
| DeepSeek-V4.1 Flash | LLM | 552B | 0.15$ / 0.60$ |
| Gemma-4 31B | LLM | 31B | 0.10$ / 0.34$ |
| Jev 1.14 | TDM | N/A | 0.042$ per 1M input tokens |
| GPT-6 Luna Decisions | TDM | N/A | 0.10$ per 1M input tokens |
| TabPFN v2 | TFM | 7.24M | Local |
| TabFM | TFM | 1.64B | Local |
| XGBoost | Baseline | N/A | Local |

*N/A marks a model for which a parameter count is not published or not applicable.*

All models, except XGBoost, were used without fine-tuning. For each prediction, the models received the training examples as context together with the test sample. The local TFMs were similarly provided with the training data as in-context examples and did not update their pretrained weights. In contrast, XGBoost was fitted to the training data for each replicate and therefore represents a conventional supervised learning approach.

The models use different input representations. The LLMs and TDMs received peptide sequences as text. For the surface activity task, pH and tail length were provided as additional input variables. TabPFN[23] and TabFM[24] received numerically encoded peptide sequences, with each residue represented by a positional token. XGBoost used a one-hot representation of the same positional sequence encoding. The additional surface activity variables were provided as numerical features to the local models.

The sequence encoder preserves the original sequence representation, including residue case. Sequences were padded to the maximum sequence length observed for each task. The model identifiers, provider information, API settings, computational details, complete encoding procedure and model-specific input formats are described in Section S1.2.

## 2.3 In-context learning protocol

For each task, models received a labeled reference set drawn from the training data and were evaluated on the same fixed test set. We used either the full or the class-balanced training set as the reference set. In standard in-context learning, the reference set contained the correct training labels. In the zero-shot condition, no labeled reference examples were provided. The same zero-shot predictions were used for comparison with both reference data variants.

For the LLMs, each request contained task instructions, a description of the prediction task, the allowed class labels, the labeled reference examples, the target samples, and a prescribed JSON output schema. The models returned class predictions and a brief reasoning summary for each batch. These summaries were mainly enforced to provide additional context for the LLM-based decisions, but may not be fully accurate[31,32] and were not used for evaluation. The TDMs received the task instructions and labeled reference examples in a structured state, followed by a separate classification question for each target sample.

To determine if the models perform better when predicting in batches or individually, we evaluated two inference modes for the LLMs and TDMs.[33] In one-by-one mode, each request contained one target sample and the complete reference set. In all-at-once mode, a single request contained the complete test set and the reference examples. Before the main evaluation, we compared the two modes on the surface activity and self-assembly tasks, using both the balanced and full training sets and all remote, API-accessed endpoint models. Based on the results of this ablation, we decided to request only one outcome per request for the main evaluation.

To test whether predictions depended on the supplied labels, we also evaluated a corrupted-label condition for the endpoint models. In this condition, every reference label was changed while retaining all declared classes: binary labels were inverted, and the three-class cancer labels were cyclically shifted. Test labels remained unchanged. The corrupted reference set was generated once for each condition. The detailed relabeling procedure is described in more detail in Section 1.3.

Model predictions can vary between replicates. No seed was sent to the endpoint models, so they were evaluated using five independent API calls. Local models were evaluated using 100 replicates, with a different fixed seed for each replicate. Unless otherwise noted, reported means and standard deviations are calculated across these replicates. This design accounts for differences in reproducibility between API-based inference and local model fitting.

The complete evaluation campaign comprised 3,725 replicate evaluations. Section S1.3 provides the prompt templates, output schema, replicate design, and seed handling. Section S1.4 describes the statistical tests.

## 2.4 Evaluation and statistics

Macro-F1 was the primary evaluation metric. It was calculated as the unweighted mean of the per-class F1 scores for each task. For binary classification tasks, we also report positive-class F1, corresponding to the true class. Secondary metrics included per-class precision, recall, F1, and support, together with confusion matrices. Token use, inference cost, and latency are reported in Table S8.

For each task, reference data variant, model, and applicable inference mode, we report the mean and standard deviation of each metric across replicates. These statistics describe variability across repeated evaluations on the same fixed train–test split. Furthermore, to compare one-by-one and all-at-once inference, we calculated the paired difference in performance (one-by-one minus all-at-once) for each task, model, and reference data variant. The comparison comprised 20 pairs of matched subgroups: two tasks, five endpoint models, and two reference data variants. We report the mean paired difference, its 95% $t$ confidence interval, and two-sided paired $t$-test and Wilcoxon signed-rank test results. Section S1.4 provides the metric definitions and statistical procedures. Sections S2 and S3 contain the complete result tables and supplementary figures.

# 3. Results & Discussion

## 3.1 Predictive performance on full training sets

The main comparison in this study is the systematic comparison of various ICL methods against a classical baseline on diverse classification tasks. Figure 2 and Table S1 summarize the performance of all eight models on the five tasks using the full training sets. All endpoint models used one-by-one inference. Here, “highest-performing” denotes the model with the highest observed mean macro-F1 in a given task. Figure S1 shows the corresponding performance with the class-balanced training sets.

Performance varied substantially across tasks, and no model dominated across the benchmark. GPT-6.1 Sol achieved the highest mean macro-F1 on three tasks, whereas XGBoost and TabFM each achieved the highest mean on one task. The spread between the best and second-best mean score ranged from 0.008 to 0.048 macro-F1 on the binary tasks.

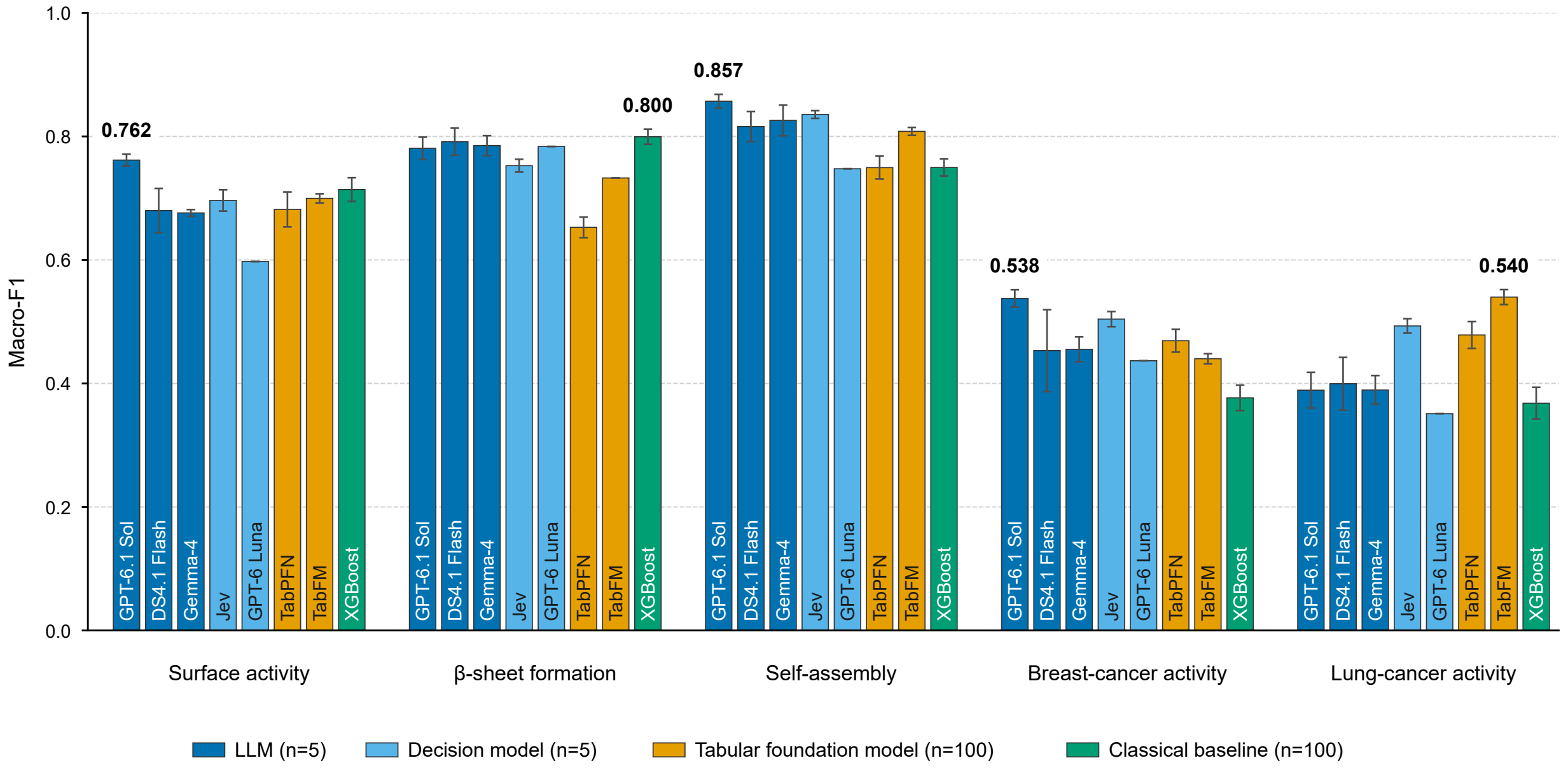

**Figure 2.** Mean macro-F1 of the eight models on the five tasks with full training sets. Colour gives the model family: LLM, decision model, tabular foundation model, or classical baseline. The legend gives the number $n$ of replicates per model family: five for the endpoint models and 100 for the local models. Endpoint models use one-by-one inference. Error bars give the standard deviation across replicates. The bold number above the best bar in each task is its mean macro-F1.

All models had lower macro-F1 on the two three-class cancer activity tasks than on the three binary tasks. On the breast cancer activity dataset, mean scores ranged from 0.377 to 0.538, with GPT-6.1 Sol performing best. On the lung cancer activity dataset, mean scores ranged between 0.351 and 0.540, with TabFM achieving the highest score. Per-class inspection showed that several models never predicted the rare very active class. On lung-cancer activity, GPT-6.1 Sol, Gemma-4 31B, GPT-6 Luna Decisions, and TabPFN had zero recall for this class, whereas on breast-cancer activity, neither TabPFN nor TabFM predicted it. These failures are only partly reflected in the aggregate macro-F1 scores and highlight the difficulty of minority-class prediction in the multiclass tasks. Table S3 provides the per-class results, and Table S4 reports precision, recall, and F1 for the very-active class. Because the breast- and lung-cancer activity tasks differ from the other, binary tasks in both class structure and class distribution, so the lower scores cannot be attributed to either factor alone.

Overall, model rankings differed substantially between tasks. XGBoost had the highest mean on beta-sheet formation but the lowest on breast-cancer activity, while TabFM had the highest mean on lung-cancer activity but the second-lowest on beta-sheet formation. To test whether these rank differences are resolvable at the present sample sizes, we bootstrapped the test partition (2,000 resamples, with peptides resampled as clusters for surface activity) and recomputed the macro-F1 difference between the highest- and second-highest-ranked model on each task (Table S10). In all five tasks this difference was not significant. Thus, the results do not support a single model ranking best across these tasks.

Replicate-to-replicate variability also differed between models and tasks. Some endpoint models had larger standard deviations than the local models, but this pattern was not consistent across all comparisons. GPT-6 Luna Decisions had no variation in macro-F1 at the displayed precision on any task, as did TabFM on beta-sheet formation. Low replicate variability for GPT-6 Luna Decisions indicates that the model is deterministic for the same input, although this was not disclosed by the model provider.

Having established a baseline performance for various foundation models against the XGBoost baseline, we next inspected how and to what extent different inference strategies affect predictive performance.

### 3.2 Inferring one-by-one versus all-at-once

Whereas deterministic classical ML baselines operate by inferring individual examples after training, the transformer-based architecture of the foundation models we tested process the full training and inference data simultaneously. As a result, inference can be done either by one-by-one prediction of individual test samples, or all-at-once inference of the full test set. We compared these two approaches to evaluate the impact of inference mode on the predictive performance. For this, we included two tasks, surface activity and self-assembly, five endpoint models, and two reference sets, full and class-balanced. This gave 20 combinations of task, model, and reference set, with both inference modes evaluated for each combination. Figure 3 shows the paired differences for each task, model, and reference set. Table S2 reports the group-level paired differences in macro-F1, and Table S5 gives the per-stratum results. Positive differences therefore indicate better performance with one-by-one inference.

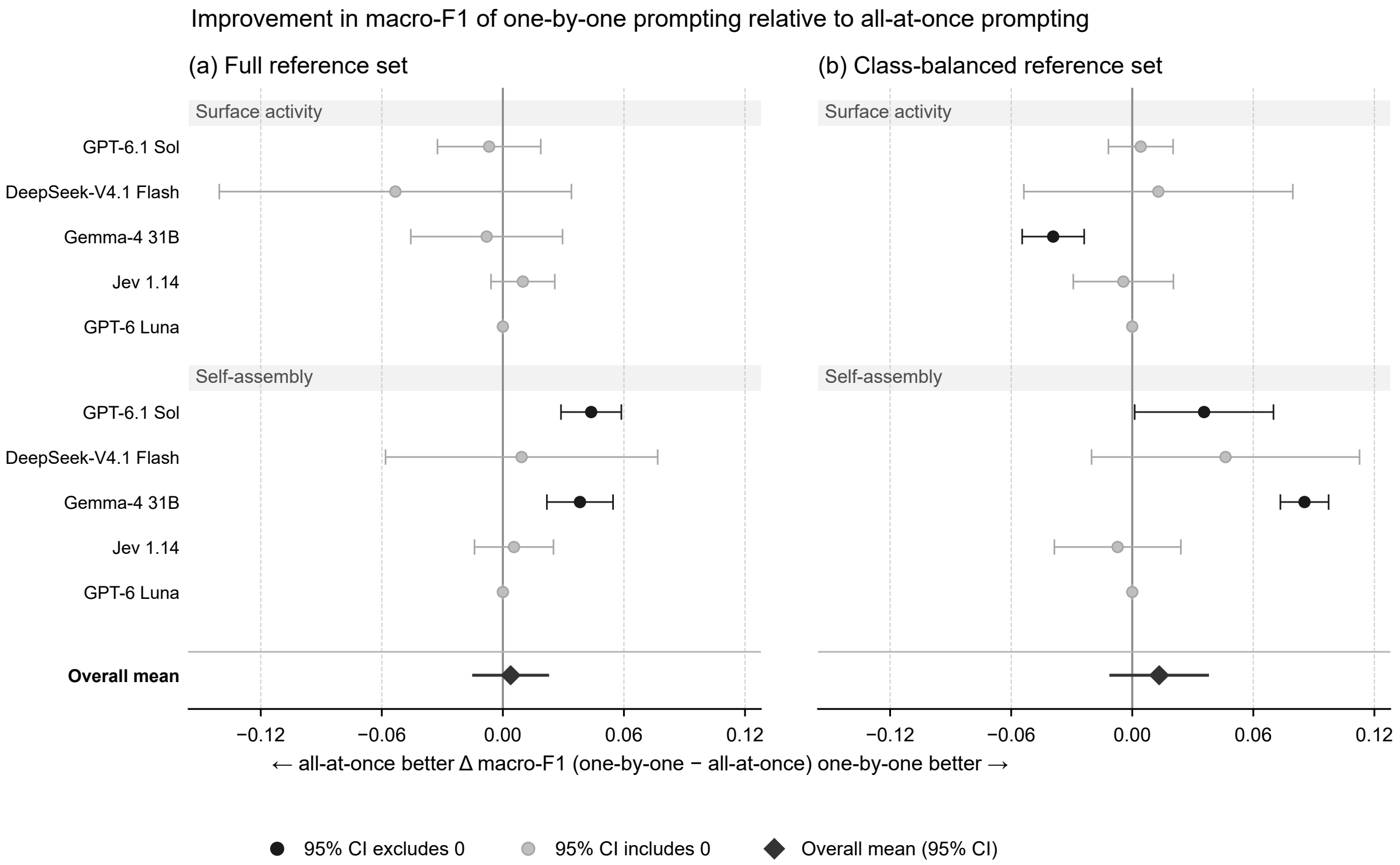

**Figure 3.** One-by-one versus all-at-once inference with the (a) full and (b) class-balanced reference sets. Each point is the mean paired difference in macro-F1 (one-by-one minus all-at-once) for one model on surface activity or self-assembly, across five paired replicates. Horizontal bars give the 95% t confidence interval for each point. Dark points mark intervals that exclude zero, and light grey points mark intervals that include zero. The diamond gives the mean difference over the ten task and model combinations in a panel, and its bar gives the 95% t confidence interval. Positive values indicate better performance with one-by-one inference.

The pilot gives 20 task, model, and reference-set combinations in total. Local models are not included. Across all 20 combinations, the overall mean difference was +0.0086 macro-F1 (95% confidence interval: −0.0056 to +0.0228). Neither the paired t-test ($p = 0.2207$) nor the Wilcoxon signed-rank test ($p = 0.2146$) showed a statistically significant overall difference between the modes. No significant difference was detected within either model family. Task-specific results differed. For self-assembly, one-by-one inference improved mean macro-F1 by 0.0257 (95% confidence interval: 0.0048 to 0.0465; $p = 0.0212$). For surface activity, the mean difference was −0.0085 (95% confidence interval: −0.0237 to +0.0068; $p = 0.2399$), with no statistically significant difference between modes. Furthermore, the comparison also showed no statistically significant difference within either model family. For the three LLMs, the mean difference was +0.0140 (95% confidence interval: −0.0106 to +0.0386; $p = 0.2352$). For the two TDMs, the mean difference was +0.0005 ($p = 0.8094$). GPT-6 Luna Decisions returned identical predictions in both modes for all four combinations of task and reference set, across five replicates per combination.

Together, these results show no consistent significant effect of inference mode. The self-assembly subset improved with one-by-one inference, whereas the overall and model-family comparisons did not. Although one-by-one inference requires more API requests, this approach is more robust against failed requests. In the case of one-by-one inference, failures only affect a single target sample rather than a complete batch.

### 3.3 Zero-shot and corrupted-label conditions

The performance comparison identifies the highest-performing models, but it does not establish whether those predictions are driven by the supplied reference examples or by information acquired during pretraining. In fact, the identical predictions returned for both inference modes for GPT-6 Luna Decisions appears to imply some models may predict from memory rather than inference. To assess the dependence of endpoint models on the provided context, we compared standard in-context learning with two conditions: a zero-shot condition, in which no labeled reference examples were supplied, and a corrupted-label condition, in which the reference labels were flipped.

Figure 4 summarizes these comparisons, where all results use the full training sets and one-by-one inference. Corresponding results for the balanced training sets are provided in Figure S2 and Table S6. Local models were not included because analogous label-corruption ablations are not applicable to conventionally trained models.

Both conditions reduced the performance of every endpoint model on every task. Relative to standard in-context learning, the decrease in macro-F1 ranged from 0.030 to 0.492 in the zero-shot condition and from 0.169 to 0.693 in the corrupted-label condition. The largest zero-shot reduction was observed for the Jev TDM on self-assembly ($\Delta = 0.492$), whereas the largest corrupted-label reduction was observed for GPT-6.1 Sol on self-assembly ($\Delta = 0.693$). This is the largest corrupted-label reduction in Figure 4.

Corrupted labels were more damaging than missing labels in 22 of the 25 model–task combinations. The mean loss across the five tasks was 0.223 in the zero-shot condition and 0.432 in the corrupted-label condition. Because corrupted-label scores were nearly constant across models and tasks (mean 0.200, range 0.131–0.284), the magnitude of the corrupted-label loss largely follows the standard-ICL baseline. The comparison between corrupted and zero-shot losses is therefore the more informative test of label dependence. The corrupted-label loss was larger on all five tasks for GPT-6.1 Sol, DeepSeek-V4.1 Flash, and Gemma-4 31B, on four tasks for Jev 1.14, and on three tasks for GPT-6 Luna Decisions. The three exceptions were GPT-6 Luna Decisions on breast-cancer activity (loss 0.273 under corrupted labels versus 0.277 under zero-shot prediction) and on lung-cancer activity (0.197 versus 0.273), and Jev 1.14 on lung-cancer activity (0.233 versus 0.239). Table S7 gives the cell values, and Appendix F.5 gives the comparison. The larger loss in performance for inference with corrupting labels than with missing labels indicates that the models generally used the supplied labels and not only the similarity between the query and the reference sequences. A model that relied only on sequence similarity would not be expected to lose more performance when the labels were corrupted than when they were removed.

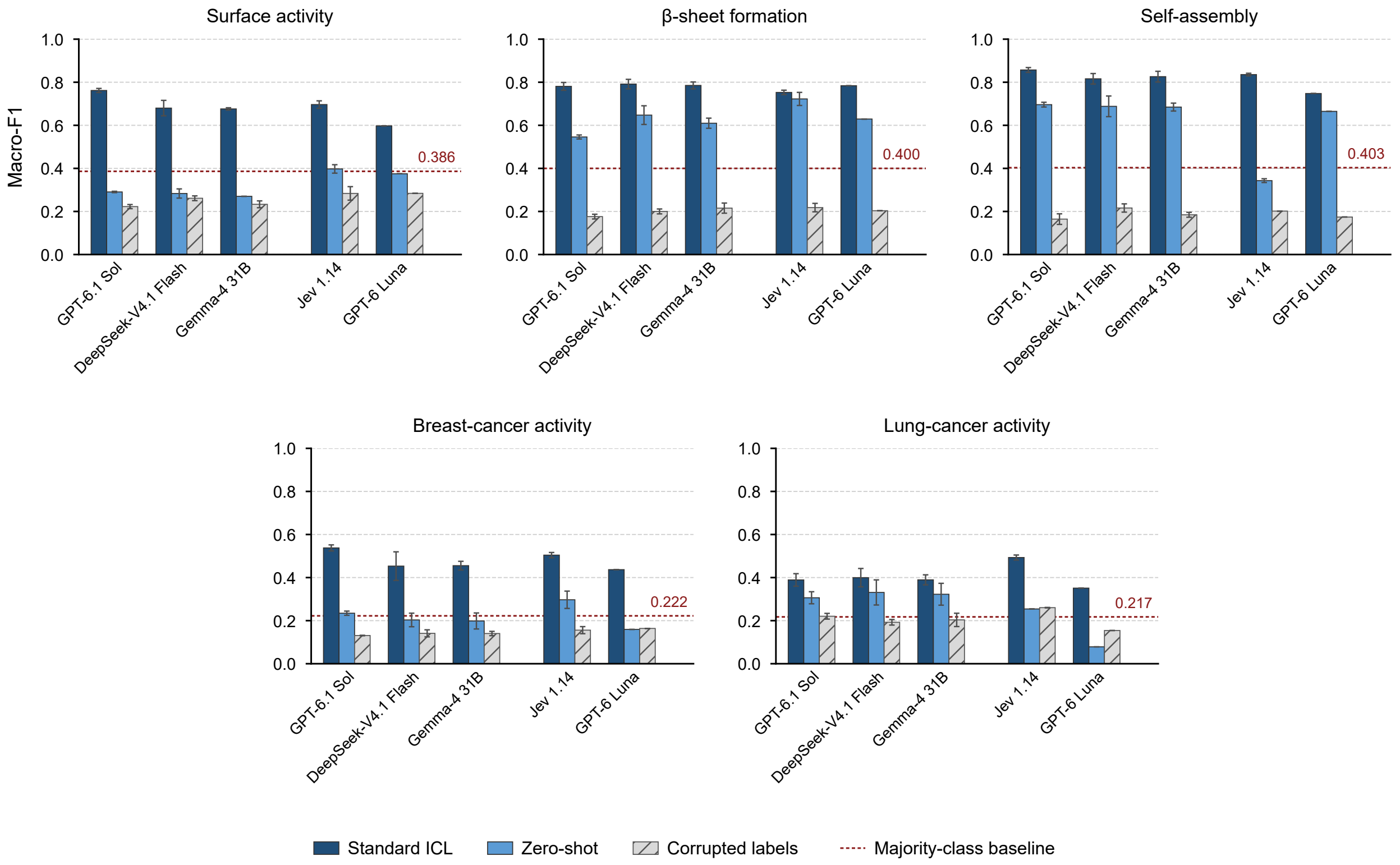

**Figure 4.** Label ablations for the five endpoint models on the five tasks. Within each model, bars give mean macro-F1 for standard in-context learning (dark blue), zero-shot prediction (light blue), and corrupted reference labels (hatched). Models are grouped by family, with LLMs before TDMs. Error bars give the standard deviation across five replicates. Dashed horizontal lines give the macro-F1 of a majority-class predictor. This predictor always selects the most frequent test label. All results use full training and one-by-one inference. Local models are not included because label ablations do not apply to them.

The three LLMs and the two decision models all showed substantial reductions in performance when labels were corrupted or missing. The overall stronger decrease when labels were corrupted is consistent with the endpoint models using the in-context reference labels when making predictions. However, the conditions do not demonstrate that the models rely exclusively on the provided context. A model may simultaneously use the supplied examples and information acquired during pretraining, including knowledge of peptide sequences, peptide properties, or related published datasets.

The interpretation is particularly informative for the surface-activity dataset, as it was collected from a recent study that was published after the training cutoffs of the evaluated endpoint models.[6] Hence, the exact dataset is unlikely to have been present during pretraining. A comparatively strong loss in performance is observed upon scrambling labels. Conversely, the loss in performance scrambled labels of sequences from datasets that were published before training cutoffs is relatively modest. This observation suggests that the LLMs may benefit from information supplied in-context, but also appears to suggest that data to which the model has been exposed during pretraining may leak into the predictions. While such reliance on knowledge that is obtained from pretraining may be advantageous when predicting or optimizing properties in low-data regimes, these effects are also likely to introduce potentially unwanted biases when searching for chemical novelty.

## Discussion

This study evaluated whether pretrained foundation models can predict peptide properties through in-context learning in low-data settings. Across five peptide classification tasks, ICL-based approaches achieved predictive performance that was generally competitive with a strong classical baseline. However, performance remained strongly task-dependent, and no model family consistently outperformed the others. GPT-6.1 Sol achieved the highest mean macro-F1 on three tasks, whereas XGBoost and TabFM each achieved the highest mean on one task. The absence of a universally best model was particularly notable given the large differences in evaluation cost. Endpoint-model API costs, including retries, varied by more than an order of magnitude across the benchmark, from USD 0.97 for Jev 1.14 to USD 27.74 for DeepSeek-V4.1 Flash, yet higher cost did not consistently translate into better predictive performance. Jev 1.14 achieved a higher mean macro-F1 across the five tasks than DeepSeek-V4.1 Flash and Gemma-4 31B (USD 11.80), and the highest mean was achieved by GPT-6.1 Sol (USD 27.62), the second most expensive endpoint model. Table S8 provides additional information regarding costs and token usage for endpoint models.

A notable finding is that tabular foundation models achieved performance comparable to both classical machine-learning models and language-based foundation models despite the absence of peptide-specific pretraining. TabFM achieved the highest performance on the lung-cancer activity task, and both TFMs remained competitive across multiple datasets. These results suggest that strong statistical priors learned during large-scale pretraining on tabular data can be sufficient for many low-data peptide prediction problems.

The inference-mode test showed no significant overall difference between one-by-one and all-at-once prediction. Nevertheless, one-by-one inference was selected for the main evaluation because it improved performance on the self-assembly task and limited the impact of a failed request to a single target sample. Because the inference comparison included only two tasks, this choice should be viewed as a practical decision rather than evidence that one inference mode is generally superior.

Overall, the results indicate that foundation-model-based ICL is a viable alternative to conventional machine-learning workflows for classification tasks in peptide science, where datasets frequently contain only tens to hundreds of labelled sequences. The principal advantage of ICL is the reduced barrier to model deployment. Although ICL still requires feature engineering, it avoids hyperparameter optimization and training, lowering the practical burden of deployment compared to conventional machine learning. While our benchmark suggest slightly better average performance for ICL methods, performance varied substantially across tasks, and the strong drop in zero-shot performance on the surface-tension dataset suggests that pretraining data leakage may inflate language-model performance. Consequently, classical baseline models, such as XGBoost remained highly competitive.

## Limitations and future directions

Several limitations should be considered when interpreting these results. First, the benchmark comprises five peptide classification tasks and a single train-test split for each task. The reported replicate statistics therefore quantify replicate-to-replicate variability on a fixed split rather than variation across independent dataset partitions. Second, the endpoint models were evaluated through external APIs. These models can change over time, may not provide deterministic outputs, and cannot always be reproduced exactly.[34] Although repeated evaluations were used to estimate variability, exact replication may not be possible in the future. Third, the comparison between one-by-one and all-at-once inference was based on only two tasks and was subsequently used to define the inference mode for the main evaluation. The inference-mode results should therefore be regarded as exploratory. Fourth, the benchmark evaluates classification tasks only. The conclusions may not be generalized to regression problems, generative peptide design, ranking tasks, or optimization settings that are common in peptide discovery workflows. Finally, this study evaluates retrospective prediction only. It does not assess active learning, iterative experimental campaigns, uncertainty-guided acquisition, or prospective prediction on newly generated peptide datasets. These applications are particularly relevant for peptide discovery and materials development and represent important directions for future work.

The datasets used here also cover a limited region of peptide chemical space and predominantly sequence-based representations. Additional benchmarks containing broader peptide chemistries, richer molecular descriptors, and prospectively collected experimental data will be required to determine how broadly the present conclusions generalize.

## Conclusions

We systematically evaluated in-context learning for peptide property prediction across five low-data classification tasks spanning self-assembly, surface activity, spectroscopy-derived structural descriptors, and biological activity. Large language models, typed decision models, and tabular foundation models all achieved predictive performance that was competitive with or outperformed an XGBoost classical baseline, demonstrating that useful peptide property prediction is possible without task-specific model training.

The zero-shot and corrupted-label conditions showed that all endpoint models benefited from the supplied reference examples. Corrupted labels produced larger reductions than the removal of examples in 22 of the 25 model–task combinations, which is consistent with the models using the supplied labels rather than only sequence similarity or pretrained knowledge. These analyses cannot exclude contributions from pretraining knowledge, however, and emphasize the need for contamination-aware benchmarking and prospective validation in future studies.

Overall, our results suggest that in-context learning is a practical addition to the peptide-machine-learning toolbox rather than a replacement for conventional methods. For small peptide datasets, foundation models offer a low-barrier route to predictive modelling, but model selection should remain task-specific and should consider predictive performance, reproducibility, and operational cost.

## Data Availability

The run records are available in the Zenodo repository at 10.5281/zenodo.23834634. Each record provided in the repository gives the predictions, the confusion matrix, the macro-F1, and the per-class precision, recall, F1, and support. Tables S1–S7 can be reproduced from these records. The datasets and split files are provided with this article, and Table S9 gives the split counts. The code used to obtain the results is supplied at https://www.github.com/BigChemistry-RobotLab/icl-benchmark.

## Acknowledgements

We acknowledge funding from the National Growth Fund project “Big Chemistry” (1420578).

## Conflicts of Interest

The authors declare no conflicts of interest.

## References

1. Li Z, Ulijn RV, W. A. Swanson H. Molecular Conformation as a Determinant of Tripeptide Self-Assembly. *The Journal of Physical Chemistry B* 2026;130(42):11190–11201, DOI: 10.1021/acs.jpcb.6c03754.

2. Sheehan F, Sementa D, Jain A, Kumar M, Tayarani-Najjaran M, Kroiss D, Ulijn RV. Peptide-Based Supramolecular Systems Chemistry. *Chemical Reviews* 2021;121(22):13869–13914, DOI: 10.1021/acs.chemrev.1c00089.
3. Zozulia O, Dolan MA, Korendovych IV. Catalytic Peptide Assemblies. *Chem Soc Rev* 2018;47(10):3621–3639, DOI: 10.1039/C8CS00080H.
4. Batra R, Loeffler TD, Chan H, Srinivasan S, Cui H, Korendovych IV, Nanda V, Palmer LC, Solomon LA, Fry HC, Sankaranarayanan SKRS. Machine Learning Overcomes Human Bias in the Discovery of Self-Assembling Peptides. *Nature Chemistry* 2022;14:1427–1435, DOI: 10.1038/s41557-022-01055-3.
5. Xu T, Wang J, Zhao S, Chen D, Zhang H, Fang Y, Kong N, Zhou Z, Li W, Wang H. Accelerating the Prediction and Discovery of Peptide Hydrogels with Human-in-the-Loop. *Nature Communications* 2023;14(1):3880, DOI: 10.1038/s41467-023-39648-2.
6. Houx J, Kong MJW, Tao S, Dankloff PFJ, Martens G, Hoeksel J, Robinson WE, Huck WTS, Mabesoone MFJ. Data-Driven Discovery of pH-Dependent Surfactant Behavior in a Peptide Amphiphile Library. *ChemRxiv* 2026, DOI: 10.26434/chemrxiv.15010226.v1.
7. Talluri YN, Sankaranarayanan SK, Fry HC, Batra R. Discovery of Unconventional and Nonintuitive Self-Assembling Peptide Materials Using Experiment-Driven Machine Learning. *Sci Adv* 2025;11(24):eadt9466, DOI: 10.1126/sciadv.adt9466.
8. Njirjak M, Žužić L, Babić M, Janković P, Otović E, Kalafatovic D, Mauša G. Reshaping the Discovery of Self-Assembling Peptides with Generative AI Guided by Hybrid Deep Learning. *Nat Mach Intell* 2024;6(12):1487–1500, DOI: 10.1038/s42256-024-00928-1.
9. Dražić E, Jelušić D, Janković Bevandić P, Mauša G, Kalafatovic D. Using Machine Learning to Fast-Track Peptide Nanomaterial Discovery. *ACS Nano* 2025;19(22):20295–20320, DOI: 10.1021/acsnano.5c00670.
10. Santos-Júnior CD, Torres MDT, Duan Y, Rodríguez Del Río Á, Schmidt TSB, Chong H, Fullam A, Kuhn M, Zhu C, Houseman A, Somborski J, Vines A, Zhao X-M, Bork P, Huerta-Cepas J, De La Fuente-Nunez C, Coelho LP. Discovery of Antimicrobial Peptides in the Global Microbiome with Machine Learning. *Cell* 2024;187(14):3761-3778.e16, DOI: 10.1016/j.cell.2024.05.013.
11. Wan F, Wong F, Collins JJ, De La Fuente-Nunez C. Machine Learning for Antimicrobial Peptide Identification and Design. *Nat Rev Bioeng* 2024;2(5):392–407, DOI: 10.1038/s44222-024-00152-x.
12. Janković P, Otović E, Mauša G, Kalafatovic D. Manually Curated Dataset of Catalytic Peptides for Ester Hydrolysis. *Data in Brief* 2023;48:109290, DOI: 10.1016/j.dib.2023.109290.
13. Van Tilborg D, Brinkmann H, Criscuolo E, Rossen L, Özçelik R, Grisoni F. Deep Learning for Low-Data Drug Discovery: Hurdles and Opportunities. *Current Opinion in Structural Biology* 2024;86:102818, DOI: 10.1016/j.sbi.2024.102818.
14. Jablonka KM, Schwaller P, Ortega-Guerrero A, Smit B. Leveraging Large Language Models for Predictive Chemistry. *Nat Mach Intell* 2024;6(2):161–169, DOI: 10.1038/s42256-023-00788-1.
15. Ramos MC, Collison CJ, White AD. A Review of Large Language Models and Autonomous Agents in Chemistry, 2024, DOI: 10.48550/arXiv.2407.01603.
16. Van Herck J, Gil MV, Jablonka KM, Abrudan A, Anker AS, Asgari M, Blaiszik B, Buffo A, Choudhury L, Corminboeuf C, Daglar H, Elahi AM, Foster IT, Garcia S, Garvin M, Godin G, Good LL, Gu J, Xiao Hu N, Jin X, Junkers T, Keskin S, Knowles TPJ, Laplaza R, Lessona M, Majumdar S, Mashhadimoslem H, McIntosh RD, Moosavi SM, Mouriño B, Nerli F, Pevida C, Poudineh N, Rajabi-Kochi M, Saar KL, Hooriabad Saboor F, Sagharichiha M, Schmidt KJ, Shi J, Simone E, Svatunek D, Taddei M, Tetko I, Tolnai D, Vahdatifar S, Whitmer J, Wieland DCF, Willumeit-Römer R, Züttel A, Smit B. Assessment of Fine-Tuned Large Language Models for Real-World Chemistry and Material Science Applications. *Chem Sci* 2025;16(2):670–684, DOI: 10.1039/D4SC04401K.
17. Zheng Z, Rampal N, Inizan TJ, Borgs C, Chayes JT, Yaghi OM. Large Language Models for Reticular Chemistry. *Nat Rev Mater* 2025;10(5):369–381, DOI: 10.1038/s41578-025-00772-8.
18. Ansari M, Moosavi SM. Agent-Based Learning of Materials Datasets from the Scientific Literature. *Digital Discovery* 2024;3(12):2607–2617, DOI: 10.1039/D4DD00252K.
19. Brown TB, Mann B, Ryder N, Subbiah M, Kaplan J, Dhariwal P, Neelakantan A, Shyam P, Sastry G, Askell A, Agarwal S, Herbert-Voss A, Krueger G, Henighan T, Child R, Ramesh A, Ziegler DM, Wu J,

Winter C, Hesse C, Chen M, Sigler E, Litwin M, Gray S, Chess B, Clark J, Berner C, McCandlish S, Radford A, Sutskever I, Amodei D. Language Models Are Few-Shot Learners, 2020, DOI: 10.48550/arXiv.2005.14165.

20. Xie SM, Raghunathan A, Liang P, Ma T. An Explanation of In-Context Learning as Implicit Bayesian Inference, 2021, DOI: 10.48550/ARXIV.2111.02080.
21. Zheng Z, Rong Z, Rampal N, Borgs C, Chayes JT, Yaghi OM. A GPT-4 Reticular Chemist for Guiding MOF Discovery. *Angew Chem Int Ed* 2023;62(46):e202311983, DOI: 10.1002/anie.202311983.
22. Joe CY, Song K, Chang R. Evaluating In-Context Learning in Large Language Models for Molecular Property Regression. *J Comput Chem* 2026;47(2):e70308, DOI: 10.1002/jcc.70308.
23. Hollmann N, Müller S, Purucker L, Krishnakumar A, Körfer M, Hoo SB, Schirrmeister RT, Hutter F. Accurate Predictions on Small Data with a Tabular Foundation Model. *Nature* 2025;637(8045):319–326, DOI: 10.1038/s41586-024-08328-6.
24. Kong W, Ilan EL, Nie S, Narayan T, Sen R, Zhou Y, Fu D, Oymak S, Das A. TabFM: A Zero-Shot Foundation Model for Tabular Data, 2026, DOI: 10.48550/arXiv.2609.37959.
25. Tang L, Zheng Y. Typed Decision Models: An Early Evidence Audit and Evaluation Checklist, 2026, DOI: 10.48550/arXiv.2609.32160.
26. Deußer T, Ramamurthy R, Sparrenberg L, Sifa R. Evaluating and Benchmarking the System One Model Jev, 2026, DOI: 10.48550/arXiv.2609.37647.
27. Liu Y, Ding S, Zhou S, Fan W, Tan Q. MolecularGPT: Open Large Language Model (LLM) for Few-Shot Molecular Property Prediction, 2024, DOI: 10.48550/arXiv.2406.12950.
28. Chen W, Tian Y, Liao N, Deng Y, Jiang D, Cao D. TabPFN Opens New Avenues for Small-Data Tabular Learning in Drug Discovery. *J Chem Inf Model* 2026;66(7):3525–3539, DOI: 10.1021/acs.jcim.5c02823.
29. Busch M, Tacke M, Lamaka SV, Zheludkevich ML, Cyron CJ, Feiler C, Aydin RC. In-Context Molecular Property Prediction with LLMs: A Blinding Study on Memorization and Knowledge Conflicts, 2026, DOI: 10.48550/arXiv.2603.25857.
30. Grisoni F, Neuhaus CS, Gabernet G, Müller AT, Hiss JA, Schneider G. Designing Anticancer Peptides by Constructive Machine Learning. *ChemMedChem* 2018;13(13):1300–1302, DOI: 10.1002/cmdc.201800204.
31. Turpin M, Michael J, Perez E, Bowman SR. Language Models Don’t Always Say What They Think: Unfaithful Explanations in Chain-of-Thought Prompting, 2023, DOI: 10.48550/arXiv.2305.04388.
32. Chen Y, Benton J, Radhakrishnan A, Uesato J, Denison C, Schulman J, Somani A, Hase P, Wagner M, Roger F, Mikulik V, Bowman SR, Leike J, Kaplan J, Perez E. Reasoning Models Don’t Always Say What They Think, 2025, DOI: 10.48550/arXiv.2505.05410.
33. Sclar M, Choi Y, Tsvetkov Y, Suhr A. Quantifying Language Models’ Sensitivity to Spurious Features in Prompt Design or: How I Learned to Start Worrying about Prompt Formatting, 2024, DOI: 10.48550/arXiv.2310.11324.
34. Chen L, Zaharia M, Zou J. How Is ChatGPT’s Behavior Changing Over Time? *Harvard Data Science Review* 2024;6(2), DOI: 10.1162/99608f92.5317da47.

# Supporting Information

## A Systematic Comparison of In-Context Learning Strategies for Peptide Property Prediction in Low-Data Regimes

Jiantao Zhao[1*], Andrei Voinea[1*], Mathijs Mabesoone[1†]

[1]Institute for Molecules and Materials, Radboud University, Heyendaalseweg 135, 6525AJ Nijmegen, The Netherlands

## S1. Supplementary methods

### S1.1 Data preparation and split construction

We used five peptide-classification tasks: surface activity[1], $\beta$-sheet formation[2], self-assembly[3], breast-cancer activity, and lung-cancer activity[4]. For surface activity, we grouped 4,233 measurements from 154 peptide sequences by sequence and pH. We kept the measurement with the lowest surface tension in each group. When values tied, we kept the first row in the source file. The resulting dataset contained 409 sequence-pH examples. All peptides had a C16 tail, and measurements were retained at pH 3, 7, and 11. The model inputs included sequence, pH, and tail length. We excluded concentration, solvent, purity, instrument metadata, and experiment identifiers. We classified a peptide as surface-active when its surface tension was below 50 mN/m.

For $\beta$-sheet formation and self-assembly, we used the source examples and labels. We mapped $\beta$-sheet classes 0.0 and 1.0 to false and true, respectively, and self-assembly classes 0 and 1 to false and true. For the breast- and lung-cancer tasks, we retained only experimentally validated entries from the Grisoni studies and kept the original labels: experimentally inactive, moderately active, and very active.

The published Houx et al. dataset includes 16 additional PA sequences that were not in the dataset used here: SKR, SRI, HSH, KPE, QEH, RYK, RQA, QAH, DRL, DHK, KSH, SQH, LEH, NKR, KDD, and QSH. This broader scope does not affect the conclusions of this work.

Case and modification notation were kept as in the source tables: lowercase residues occur in two breast-cancer sequences, and four self-assembly sequences contain the modified-residue notation `P'`. The source tables do not provide a complete modification dictionary, so no chemical meaning was assigned to unverified notation. No sequence normalization or deduplication was applied. No sequence is repeated in the four non-surface tables, so their row-level splits also separate sequences.

The four non-surface tasks used a stratified 80:20 split with seed 42. The surface-activity split used a shuffled, stratified group five-fold split by peptide sequence. The fold that minimized the sum of the absolute test-fraction deviation from 20% and the absolute class-proportion deviations from the cleaned dataset was selected, and ties went to the lower fold number. Class-balanced subsets were drawn from the full training rows with the same fixed seed and contained an equal number of rows per class. Table S9 gives cleaned and per-class counts, the split sizes, and the balanced training sizes.

### S1.2 Models and input representations

Eight models from four families were compared. The endpoint models were three large language models (LLMs; GPT-6.1 Sol, DeepSeek-V4.1 Flash, and Gemma-4 31B) and two typed decision models (TDMs; Jev 1.14 and GPT-6 Luna Decisions). The local models were two tabular foundation models (TabPFN v2 and TabFM) and one gradient-boosted tree baseline (XGBoost). We selected TabPFN v2 instead of its newer versions due to licensing restrictions and access gating on the newer models.

Endpoint models were accessed through OpenRouter with default provider-selected hyperparameters. DeepSeek-V4.1 Flash used low reasoning effort, while the other models used their provider default reasoning effort. The decision endpoints did not accept a temperature or seed. The tabular foundation models ran on one machine with an NVIDIA RTX 4090 GPU and 64 GB of RAM, while XGBoost used the CPU.

Endpoint models received peptide sequences as text. For surface activity, the prompt also included pH and tail length. TabPFN v2 and TabFM received positional integer tokens, with pH and tail length supplied as numeric features for surface activity, while XGBoost received positional one-hot tokens and the same numeric features. The encoder used a 43-token vocabulary, preserved case and modification notation such as P', and padded sequences to the task maximum length (3, 5, 24, 38, and 38 residues). Padding tokens were encoded separately. TabPFN v2 used its default classifier with categorical feature indices from 0 to the task maximum length minus 1. TabFM used its default row and feature limits. XGBoost used the library defaults with 0.8 row and 0.8 column subsampling and the replicate seed as the random state.

Endpoint models used one-by-one inference. TabPFN v2 and TabFM used the training rows as in-context data without weight updates. XGBoost was fitted to the training rows in each replicate. All local models predicted the same fixed test rows. The costs in Table S8 are provider-reported totals over all requests.

### S1.3 In-context learning protocol and label controls

Each LLM prompt contained the task instruction, the labeled reference set, the target peptides, the allowed labels, and a JSON output schema. The schema required one label per target and a short reasoning field. The TDMs received the reference rows in a structured state and one choice question per target. No retries were allowed during our requests, and each failed request was repeated in a new API call until a valid response was received. In one-by-one mode, each target peptide was submitted in a separate request that repeated the full reference set. In all-at-once mode, all target peptides were submitted in one request, subject to the TDM limit of 200 questions per request. Zero-shot prompts omitted the reference rows. The main evaluation used one-by-one inference. An example request and response for surface activity is described below. The prompt for the other tasks was similar, with the task instruction and allowed labels changed to match the task.

**Example prompt and response (surface activity)**

**Large Language Model**

**System message**

Predict the outputs for every target row, in the same order. Use the reference rows to infer the input to output mapping. If no reference rows are provided, use the task context for zero-shot prediction. Follow all task-specific output labels, units, formats, headers, and schema. Treat empty cells as missing values, not zero. Missing reference outputs are not examples of valid labels or predictions. Return exactly one prediction per target row in the required structured format. After the predictions, briefly state the basis for the predictions and any relevant uncertainty. Use task context and relevant general knowledge to interpret the inputs. When reference rows are available, prefer patterns supported by the examples. State relevant uncertainty. Do not invent evidence or claim validation that was not performed.

**User message**

**Context:** Predict the surface activity from the supplied examples. Every candidate is a peptide amphiphile with an N-terminal C16 tail and a tripeptide headgroup. Measurements at pH 3 use citrate buffer, measurements at pH 7 use phosphate buffer, and measurements at pH 11 use Tris buffer. A peptide is considered surface active if it has a surface tension value below 50 mN/m. Allowed output labels: `{"surface_active": ["false", "true"]}`.

**Reference data:**

```
Tail length,Peptide sequence,pH,Surface active
16,SAH,11,false
...
```

Predict “Surface active” for the following targets in row order. Use each “target_id” value as its key in the predictions object:

```
target_id,Tail length,Peptide sequence,pH
target_0,16,GRH,7
...
```

**Response format**

```
{
  "predictions": {
    "[target]": {"[label]": "[predicted class]"},
    ...
  },
  "reasoning": "[basis and uncertainty]"
}
```

### Typed Decision Model

**State**

```
[
  {
    "role": "user",
    "content": [
      {
        "type": "input_text",
        "text": "Use the labeled reference rows to infer how the
          task inputs map to the output. If no reference rows are
          provided, use the task context for zero-shot prediction.
          Empty cells are missing values, not zero.
          \nTask context:\nPredict the surface activity from the
          supplied examples. Every candidate is a peptide
          amphiphile with an N-terminal C16 tail and a tripeptide
          headgroup. Measurements at pH 3 use citrate buffer,
          measurements at pH 7 use phosphate buffer, and
          measurements at pH 11 use Tris buffer. A peptide is
          considered surface active if it has a surface tension
          value below 50 mN/m."
      },
      {
        "type": "input_json",
        "data": {
          "references": {
            "columns": ["Tail length", "Peptide sequence", "pH",
                        "Surface active"],
            "rows": [["16", "SAH", "11", "false"], "..."]
          }
        }
      }
    ]
  }
]
```

**Questions**

```
{
  "target_0": {
    "type": "choice",
    "instructions": "Is the peptide C16-GRH surface active at pH 7?",
    "criteria": {"false": "false", "true": "true"}
  },
  "...": "..."
}
```

Long string values in the state block are wrapped for display only.

Reference sets were the full training set, the class-balanced training set, a corrupted version of either set, or empty for zero-shot. In the binary tasks, corruption inverted every reference label. In the three-class cancer tasks, corruption shifted every reference label to the next class (inactive → moderately active → very active → inactive). Every reference label changed, every class remained present in the corrupted set, and test labels were unchanged. Balanced training sets were formed by undersampling each training class to the smallest training class.

Endpoint models were run in five independent repeats, and no seed was used. Local models were run in 100 repeats with seeds 42-141. Test rows and labels were fixed across models, reference sets, and repeats. The zero-shot results were shared by the full and balanced comparisons. The complete evaluation comprised 3,725 replicate evaluations.

### S1.4 Evaluation and statistics

Macro-F1 was the unweighted mean of the per-class F1 over the declared classes. Positive-class F1 was the F1 of the true class in the binary tasks (this was not defined for the three-class cancer tasks). Precision, recall, F1, and test support were calculated for each class. Unless stated otherwise, values are means over repeats, and SD is the sample standard deviation.

The one-by-one and all-at-once comparison used 20 matched strata: two tasks (surface activity and self-assembly), five endpoint models, and two reference sets (full and class-balanced). For Table S5, the five replicate-level paired differences in each stratum gave a mean difference and a 95% t interval. The difference was one-by-one minus all-at-once. For Table S2, replicate scores were averaged within each stratum before the 20 matched differences were formed. The summary tests therefore used n = 20 paired differences. These differences were tested with a two-sided paired t-test and a two-sided Wilcoxon signed-rank test. The tests were exploratory, and no correction for multiple comparisons was applied.

Label-control losses were standard ICL minus the condition. The zero-shot loss used the standard full-training score. The corrupted-full and corrupted-balanced losses used the matching standard score. A positive loss indicates that the condition reduced macro-F1.

## S2. Supplementary tables

**Table S1.** Full-training macro-F1 (mean ± SD) for the eight models on the five tasks. Endpoint models used one-by-one inference (n = 5); local models used their training rows directly (n = 100).

| Model | Family | Surface activity | $\beta$-sheet formation | Self-assembly | Breast-cancer activity | Lung-cancer activity |
|---|---|---|---|---|---|---|
| GPT-6.1 Sol | LLM | 0.762 ± 0.010 | 0.781 ± 0.018 | 0.857 ± 0.011 | 0.538 ± 0.014 | 0.389 ± 0.029 |
| DeepSeek-V4.1 Flash | LLM | 0.680 ± 0.036 | 0.791 ± 0.022 | 0.816 ± 0.024 | 0.453 ± 0.066 | 0.400 ± 0.043 |
| Gemma-4 31B | LLM | 0.676 ± 0.005 | 0.785 ± 0.016 | 0.826 ± 0.025 | 0.455 ± 0.020 | 0.389 ± 0.023 |
| Jev 1.14 | TDM | 0.696 ± 0.017 | 0.753 ± 0.010 | 0.836 ± 0.006 | 0.504 ± 0.012 | 0.493 ± 0.012 |
| GPT-6 Luna Decisions | TDM | 0.597 ± 0.000 | 0.784 ± 0.000 | 0.748 ± 0.000 | 0.437 ± 0.000 | 0.351 ± 0.000 |
| TabPFN v2 | Tabular foundation model | 0.682 ± 0.028 | 0.653 ± 0.017 | 0.750 ± 0.019 | 0.469 ± 0.018 | 0.479 ± 0.022 |
| TabFM | Tabular foundation model | 0.700 ± 0.007 | 0.733 ± 0.000 | 0.808 ± 0.006 | 0.440 ± 0.008 | 0.540 ± 0.012 |
| XGBoost | Classical baseline | 0.714 ± 0.019 | 0.800 ± 0.012 | 0.750 ± 0.014 | 0.377 ± 0.021 | 0.368 ± 0.026 |

**Table S2.** Summary of the inference mode comparison. Mean differences are macro-F1 (one-by-one minus all-at-once); intervals are 95% t intervals. The tests were two-sided and exploratory.

| Group | n | Mean difference | 95% t interval | Paired t p | Wilcoxon p |
|---|---|---|---|---|---|
| All endpoint models | 20 | +0.0086 | −0.0056 to +0.0228 | 0.2207 | 0.2146 |
| LLMs | 12 | +0.0140 | −0.0106 to +0.0386 | 0.2352 | 0.2661 |
| TDMs | 8 | +0.0005 | −0.0040 to +0.0049 | 0.8094 | 0.8750 |
| Surface activity | 10 | −0.0085 | −0.0237 to +0.0068 | 0.2399 | 0.4609 |
| Self-assembly | 10 | +0.0257 | +0.0048 to +0.0465 | 0.0212 | 0.0234 |

**Table S3.** Per-class F1 for full training. Binary classes are shown by their source labels, false and true. Values are means over n = 5 independent repeats for endpoint models and n = 100 seeded repeats for local models. Test support for each class is given in Table S9. Precision and recall are given in the deposited per-class metrics, and Table S4 gives precision, recall, and F1 for the very active class under full and class-balanced training.

| Class | GPT-6.1 Sol | DeepSeek-V4.1 Flash | Gemma-4 31B | Jev 1.14 | GPT-6 Luna Decisions | TabPFN v2 | TabFM | XGBoost |
|---|---|---|---|---|---|---|---|---|
| Surface activity, false | 0.800 | 0.733 | 0.687 | 0.791 | 0.687 | 0.824 | 0.794 | 0.769 |
| Surface activity, true | 0.724 | 0.627 | 0.665 | 0.602 | 0.508 | 0.539 | 0.606 | 0.659 |
| $\beta$-sheet formation, false | 0.851 | 0.849 | 0.825 | 0.819 | 0.836 | 0.793 | 0.817 | 0.869 |
| $\beta$-sheet formation, true | 0.710 | 0.734 | 0.745 | 0.687 | 0.732 | 0.513 | 0.649 | 0.730 |
| Self-assembly, false | 0.802 | 0.739 | 0.763 | 0.769 | 0.652 | 0.664 | 0.731 | 0.647 |
| Self-assembly, true | 0.912 | 0.893 | 0.889 | 0.902 | 0.843 | 0.835 | 0.886 | 0.853 |
| Breast-cancer activity, inactive | 0.618 | 0.485 | 0.530 | 0.592 | 0.357 | 0.702 | 0.637 | 0.592 |
| Breast-cancer activity, moderately active | 0.648 | 0.617 | 0.613 | 0.616 | 0.553 | 0.705 | 0.683 | 0.531 |
| Breast-cancer activity, very active | 0.347 | 0.258 | 0.222 | 0.305 | 0.400 | 0.000 | 0.000 | 0.007 |
| Lung-cancer activity, inactive | 0.535 | 0.415 | 0.552 | 0.522 | 0.353 | 0.764 | 0.615 | 0.474 |
| Lung-cancer activity, moderately active | 0.632 | 0.683 | 0.616 | 0.679 | 0.700 | 0.672 | 0.671 | 0.627 |
| Lung-cancer activity, very active | 0.000 | 0.100 | 0.000 | 0.279 | 0.000 | 0.000 | 0.333 | 0.003 |

**Table S4.** Precision, recall, and F1 for the very active class under full and class-balanced training on the two cancer tasks. Values are means over n = 5 (endpoint) or n = 100 (local) repeats.

| Task | Model | Full precision | Full recall | Full F1 | Balanced precision | Balanced recall | Balanced F1 |
|---|---|---|---|---|---|---|---|
| Breast-cancer activity | GPT-6.1 Sol | 0.367 | 0.333 | 0.347 | 0.175 | 0.667 | 0.276 |
| | DeepSeek-V4.1 Flash | 0.190 | 0.400 | 0.258 | 0.114 | 0.667 | 0.194 |
| | Gemma-4 31B | 0.167 | 0.333 | 0.222 | 0.139 | 0.667 | 0.229 |
| | Jev 1.14 | 0.283 | 0.333 | 0.305 | 0.193 | 0.667 | 0.299 |
| | GPT-6 Luna Decisions | 0.500 | 0.333 | 0.400 | 0.167 | 0.333 | 0.222 |
| | TabPFN v2 | 0.000 | 0.000 | 0.000 | 0.150 | 0.667 | 0.244 |
| | TabFM | 0.000 | 0.000 | 0.000 | 0.301 | 0.667 | 0.414 |
| | XGBoost | 0.007 | 0.007 | 0.007 | 0.192 | 0.460 | 0.268 |
| Lung-cancer activity | GPT-6.1 Sol | 0.000 | 0.000 | 0.000 | 0.533 | 0.360 | 0.424 |
| | DeepSeek-V4.1 Flash | 0.133 | 0.080 | 0.100 | 0.225 | 0.480 | 0.304 |
| | Gemma-4 31B | 0.000 | 0.000 | 0.000 | 0.351 | 0.480 | 0.404 |
| | Jev 1.14 | 0.467 | 0.200 | 0.279 | 0.313 | 0.600 | 0.411 |
| | GPT-6 Luna Decisions | 0.000 | 0.000 | 0.000 | 0.000 | 0.000 | 0.000 |
| | TabPFN v2 | 0.000 | 0.000 | 0.000 | 0.507 | 0.400 | 0.444 |
| | TabFM | 1.000 | 0.200 | 0.333 | 0.167 | 0.200 | 0.182 |
| | XGBoost | 0.005 | 0.002 | 0.003 | 0.143 | 0.222 | 0.173 |

When a model had no true positives for the very-active class, precision and F1 are reported as 0.000. Under full training on lung cancer, four models had zero recall for the very active class: GPT-6.1 Sol, Gemma-4 31B, GPT-6 Luna Decisions, and TabPFN v2. GPT-6.1 Sol, Gemma-4 31B, and TabPFN v2 emitted the class only as false positives; GPT-6 Luna made no very-active prediction.

**Table S5.** Mode comparison by task, reference set, and model. Values are mean macro-F1; the difference is one-by-one minus all-at-once; intervals are 95% t intervals from the five replicate-level paired differences. Differences and intervals were calculated from unrounded values.

| Task | Reference set | Model | One-by-one | All-at-once | Difference | 95% t interval |
|---|---|---|---|---|---|---|
| Surface activity | Full | GPT-6.1 Sol | 0.762 | 0.769 | −0.007 | −0.0324 to +0.0188 |
| | | DeepSeek-V4.1 Flash | 0.680 | 0.733 | −0.053 | −0.1405 to +0.0340 |
| | | Gemma-4 31B | 0.676 | 0.684 | −0.008 | −0.0456 to +0.0296 |
| | | Jev 1.14 | 0.696 | 0.686 | +0.010 | −0.0059 to +0.0257 |
| | | GPT-6 Luna Decisions | 0.597 | 0.597 | 0.000 | +0.0000 to +0.0000 |
| | Balanced | GPT-6.1 Sol | 0.727 | 0.723 | +0.004 | −0.0118 to +0.0202 |
| | | DeepSeek-V4.1 Flash | 0.672 | 0.659 | +0.013 | −0.0537 to +0.0796 |
| | | Gemma-4 31B | 0.603 | 0.642 | −0.039 | −0.0546 to −0.0238 |
| | | Jev 1.14 | 0.714 | 0.718 | −0.004 | −0.0292 to +0.0204 |
| | | GPT-6 Luna Decisions | 0.634 | 0.634 | 0.000 | +0.0000 to +0.0000 |
| Self-assembly | Full | GPT-6.1 Sol | 0.857 | 0.813 | +0.044 | +0.0288 to +0.0587 |
| | | DeepSeek-V4.1 Flash | 0.816 | 0.807 | +0.009 | −0.0582 to +0.0767 |
| | | Gemma-4 31B | 0.826 | 0.788 | +0.038 | +0.0218 to +0.0546 |
| | | Jev 1.14 | 0.836 | 0.830 | +0.006 | −0.0141 to +0.0251 |
| | | GPT-6 Luna Decisions | 0.748 | 0.748 | 0.000 | +0.0000 to +0.0000 |
| | Balanced | GPT-6.1 Sol | 0.865 | 0.830 | +0.036 | +0.0012 to +0.0700 |
| | | DeepSeek-V4.1 Flash | 0.836 | 0.790 | +0.046 | −0.0202 to +0.1126 |
| | | Gemma-4 31B | 0.851 | 0.766 | +0.085 | +0.0734 to +0.0973 |
| | | Jev 1.14 | 0.809 | 0.817 | −0.007 | −0.0386 to +0.0241 |
| | | GPT-6 Luna Decisions | 0.659 | 0.659 | 0.000 | +0.0000 to +0.0000 |

**Table S6.** Label-control macro-F1 (mean (SD)) for the five endpoint models on the five tasks. Standard full and standard balanced use the corresponding reference set. Zero-shot uses no reference rows and is shared by the full and balanced comparisons. Corrupted-full and corrupted-balanced labels changed every reference label; test labels were unchanged. n = 5 independent repeats.

| Task | Model | Standard full | Zero-shot | Corrupted full | Standard balanced | Corrupted balanced |
|---|---|---|---|---|---|---|
| Surface activity | GPT-6.1 Sol | 0.762 (0.010) | 0.291 (0.003) | 0.223 (0.010) | 0.727 (0.020) | 0.251 (0.012) |
| | DeepSeek-V4.1 Flash | 0.680 (0.036) | 0.283 (0.021) | 0.261 (0.011) | 0.672 (0.024) | 0.265 (0.041) |
| | Gemma-4 31B | 0.676 (0.005) | 0.270 (0.000) | 0.233 (0.016) | 0.603 (0.009) | 0.270 (0.017) |
| | Jev 1.14 | 0.696 (0.017) | 0.397 (0.020) | 0.284 (0.031) | 0.714 (0.015) | 0.223 (0.009) |
| | GPT-6 Luna Decisions | 0.597 (0.000) | 0.375 (0.000) | 0.284 (0.000) | 0.634 (0.000) | 0.361 (0.000) |
| $\beta$-sheet formation | GPT-6.1 Sol | 0.781 (0.018) | 0.546 (0.009) | 0.176 (0.011) | 0.746 (0.013) | 0.225 (0.030) |
| | DeepSeek-V4.1 Flash | 0.791 (0.022) | 0.647 (0.043) | 0.200 (0.011) | 0.774 (0.023) | 0.220 (0.027) |
| | Gemma-4 31B | 0.785 (0.016) | 0.610 (0.024) | 0.215 (0.024) | 0.763 (0.010) | 0.246 (0.009) |
| | Jev 1.14 | 0.753 (0.010) | 0.722 (0.030) | 0.218 (0.020) | 0.784 (0.010) | 0.194 (0.009) |
| | GPT-6 Luna Decisions | 0.784 (0.000) | 0.629 (0.000) | 0.203 (0.000) | 0.766 (0.000) | 0.257 (0.000) |
| Self-assembly | GPT-6.1 Sol | 0.857 (0.011) | 0.696 (0.011) | 0.164 (0.025) | 0.865 (0.014) | 0.189 (0.024) |
| | DeepSeek-V4.1 Flash | 0.816 (0.024) | 0.688 (0.048) | 0.216 (0.019) | 0.836 (0.036) | 0.254 (0.022) |
| | Gemma-4 31B | 0.826 (0.025) | 0.685 (0.018) | 0.184 (0.011) | 0.851 (0.014) | 0.223 (0.014) |
| | Jev 1.14 | 0.836 (0.006) | 0.343 (0.009) | 0.202 (0.000) | 0.809 (0.021) | 0.235 (0.015) |
| | GPT-6 Luna Decisions | 0.748 (0.000) | 0.665 (0.000) | 0.174 (0.000) | 0.659 (0.000) | 0.256 (0.000) |
| Breast-cancer activity | GPT-6.1 Sol | 0.538 (0.014) | 0.235 (0.010) | 0.131 (0.001) | 0.518 (0.017) | 0.177 (0.021) |
| | DeepSeek-V4.1 Flash | 0.453 (0.066) | 0.203 (0.032) | 0.141 (0.016) | 0.380 (0.017) | 0.175 (0.045) |
| | Gemma-4 31B | 0.455 (0.020) | 0.198 (0.037) | 0.140 (0.010) | 0.504 (0.026) | 0.223 (0.000) |
| | Jev 1.14 | 0.504 (0.012) | 0.297 (0.040) | 0.156 (0.016) | 0.506 (0.030) | 0.188 (0.000) |
| | GPT-6 Luna Decisions | 0.437 (0.000) | 0.159 (0.000) | 0.164 (0.000) | 0.505 (0.000) | 0.183 (0.000) |
| Lung-cancer activity | GPT-6.1 Sol | 0.389 (0.029) | 0.306 (0.028) | 0.220 (0.013) | 0.526 (0.030) | 0.233 (0.000) |
| | DeepSeek-V4.1 Flash | 0.400 (0.043) | 0.331 (0.058) | 0.192 (0.013) | 0.500 (0.054) | 0.196 (0.039) |
| | Gemma-4 31B | 0.389 (0.023) | 0.323 (0.051) | 0.203 (0.031) | 0.552 (0.022) | 0.266 (0.029) |
| | Jev 1.14 | 0.493 (0.012) | 0.254 (0.000) | 0.260 (0.001) | 0.480 (0.025) | 0.179 (0.002) |
| | GPT-6 Luna Decisions | 0.351 (0.000) | 0.078 (0.000) | 0.154 (0.000) | 0.426 (0.000) | 0.203 (0.000) |

**Table S7.** Label-control losses (standard ICL minus condition) for the cells in Table S6. Zero-shot loss uses the standard full score; corrupted-full and corrupted-balanced losses use the matching standard score. The final column compares the corrupted-full loss with the zero-shot loss: Yes or No.

| Task | Model | Zero-shot loss | Corrupted-full loss | Corrupted-balanced loss | Corrupted-full loss > zero-shot loss |
|---|---|---|---|---|---|
| Surface activity | GPT-6.1 Sol | 0.471 | 0.539 | 0.476 | Yes |
| | DeepSeek-V4.1 Flash | 0.397 | 0.419 | 0.407 | Yes |
| | Gemma-4 31B | 0.406 | 0.443 | 0.332 | Yes |
| | Jev 1.14 | 0.299 | 0.413 | 0.490 | Yes |
| | GPT-6 Luna Decisions | 0.222 | 0.313 | 0.273 | Yes |
| $\beta$-sheet formation | GPT-6.1 Sol | 0.235 | 0.605 | 0.521 | Yes |
| | DeepSeek-V4.1 Flash | 0.144 | 0.591 | 0.554 | Yes |
| | Gemma-4 31B | 0.175 | 0.570 | 0.517 | Yes |
| | Jev 1.14 | 0.030 | 0.535 | 0.590 | Yes |
| | GPT-6 Luna Decisions | 0.155 | 0.580 | 0.509 | Yes |
| Self-assembly | GPT-6.1 Sol | 0.161 | 0.693 | 0.677 | Yes |
| | DeepSeek-V4.1 Flash | 0.128 | 0.600 | 0.582 | Yes |
| | Gemma-4 31B | 0.141 | 0.641 | 0.628 | Yes |
| | Jev 1.14 | 0.492 | 0.634 | 0.574 | Yes |
| | GPT-6 Luna Decisions | 0.083 | 0.574 | 0.404 | Yes |
| Breast-cancer activity | GPT-6.1 Sol | 0.303 | 0.407 | 0.341 | Yes |
| | DeepSeek-V4.1 Flash | 0.250 | 0.312 | 0.205 | Yes |
| | Gemma-4 31B | 0.257 | 0.315 | 0.281 | Yes |
| | Jev 1.14 | 0.207 | 0.348 | 0.319 | Yes |
| | GPT-6 Luna Decisions | 0.277 | 0.273 | 0.322 | No |
| Lung-cancer activity | GPT-6.1 Sol | 0.083 | 0.169 | 0.292 | Yes |
| | DeepSeek-V4.1 Flash | 0.069 | 0.207 | 0.304 | Yes |
| | Gemma-4 31B | 0.067 | 0.186 | 0.286 | Yes |
| | Jev 1.14 | 0.239 | 0.233 | 0.301 | No |
| | GPT-6 Luna Decisions | 0.273 | 0.197 | 0.223 | No |

Losses were calculated from unrounded values.

**Table S8.** Endpoint cost, token use, request count, and mean run latency. Costs, token counts, and request counts are complete provider-log totals. Mean latency is for full-training one-by-one runs (25 runs per model); the total row gives the mean over all 125 runs.

| Model | Cost (USD) | Tokens | Requests | Mean latency (s) |
|---|---:|---:|---:|---:|
| GPT-6.1 Sol | 27.62 | 14,922,803 | 7,660 | 178.97 |
| DeepSeek-V4.1 Flash | 27.74 | 52,516,029 | 7,227 | 364.85 |
| Gemma-4 31B | 11.80 | 42,978,860 | 7,626 | 9.74 |
| Jev 1.14 | 0.97 | 23,427,583 | 7,026 | 3.52 |
| GPT-6 Luna Decisions | 1.31 | 13,102,984 | 7,023 | 3.57 |
| **Total** | **69.43** | **146,948,259** | **36,562** | **112.13** |

Costs are provider-reported totals over all API requests, including retries. The per-run records provided with this article cover completed runs only and therefore sum to a lower total. Totals are calculated from unrounded values.

**Table S9.** Dataset sizes, class counts, and split sizes. Binary classes are shown by their source labels, false and true. Full counts are the cleaned datasets. Train and test counts define the fixed splits; balanced train counts give the class-balanced training subsets. Per-class counts are listed in class order.

| Task | Total | Classes | Train | Test | Balanced train | Full per-class counts | Test per-class counts |
|---|---|---|---|---|---|---|---|
| Surface activity | 409 | 2 | 328 | 81 | 242 | False 258; True 151 | False 51; True 30 |
| $\beta$-sheet formation | 268 | 2 | 214 | 54 | 146 | False 177; True 91 | False 36; True 18 |
| Self-assembly | 368 | 2 | 294 | 74 | 190 | False 119; True 249 | False 24; True 50 |
| Breast-cancer activity | 199 | 3 | 159 | 40 | 45 | Inactive 83; Moderately active 98; Very active 18 | Inactive 17; Moderately active 20; Very active 3 |
| Lung-cancer activity | 151 | 3 | 120 | 31 | 57 | Inactive 52; Moderately active 75; Very active 24 | Inactive 11; Moderately active 15; Very active 5 |

**Table S10.** Statistical significance of the macro-F1 difference between the highest- and second-highest-ranked model on each task, using the full reference set. Δ is the observed difference in mean macro-F1 between the two models, where the mean is taken over replicates (the values in Table S1); positive values indicate that the top-ranked model scored higher than the runner-up. Confidence intervals were obtained by bootstrapping the test partition (2,000 resamples; peptides resampled as clusters for surface activity). On each resample, macro-F1 was recomputed for every replicate of both models, averaged over replicates (100 for XGBoost, TabPFN and TabFM; five for the endpoint models), and the difference between the two averages taken; the interval is the 2.5th–97.5th percentile of these differences. Because every replicate of a model shares one fixed test partition, a test across replicates would measure seed-to-seed stability rather than performance on unseen peptides; the intervals are therefore conditional on that split. Percentile intervals need not be symmetric about Δ, and the bootstrap mean differs from Δ by at most 0.007. The p-value is two-sided and uncorrected for multiple comparisons. No 95% confidence interval excludes zero.

| Task | Highest | Second | Δ | 95% CI |
|---|---|---|---|---|
| Surface activity | GPT-6.1 Sol | XGBoost | +0.048 | −0.033 to +0.130 |
| $\beta$-sheet formation | XGBoost | DeepSeek-V4.1 Flash | +0.008 | −0.126 to +0.134 |
| Self-assembly | GPT-6.1 Sol | Jev 1.14 | +0.022 | −0.048 to +0.095 |
| Breast-cancer activity | GPT-6.1 Sol | Jev 1.14 | +0.033 | −0.105 to +0.161 |
| Lung-cancer activity | TabFM | Jev 1.14 | +0.047 | −0.041 to +0.141 |

# S3. Supplementary figures

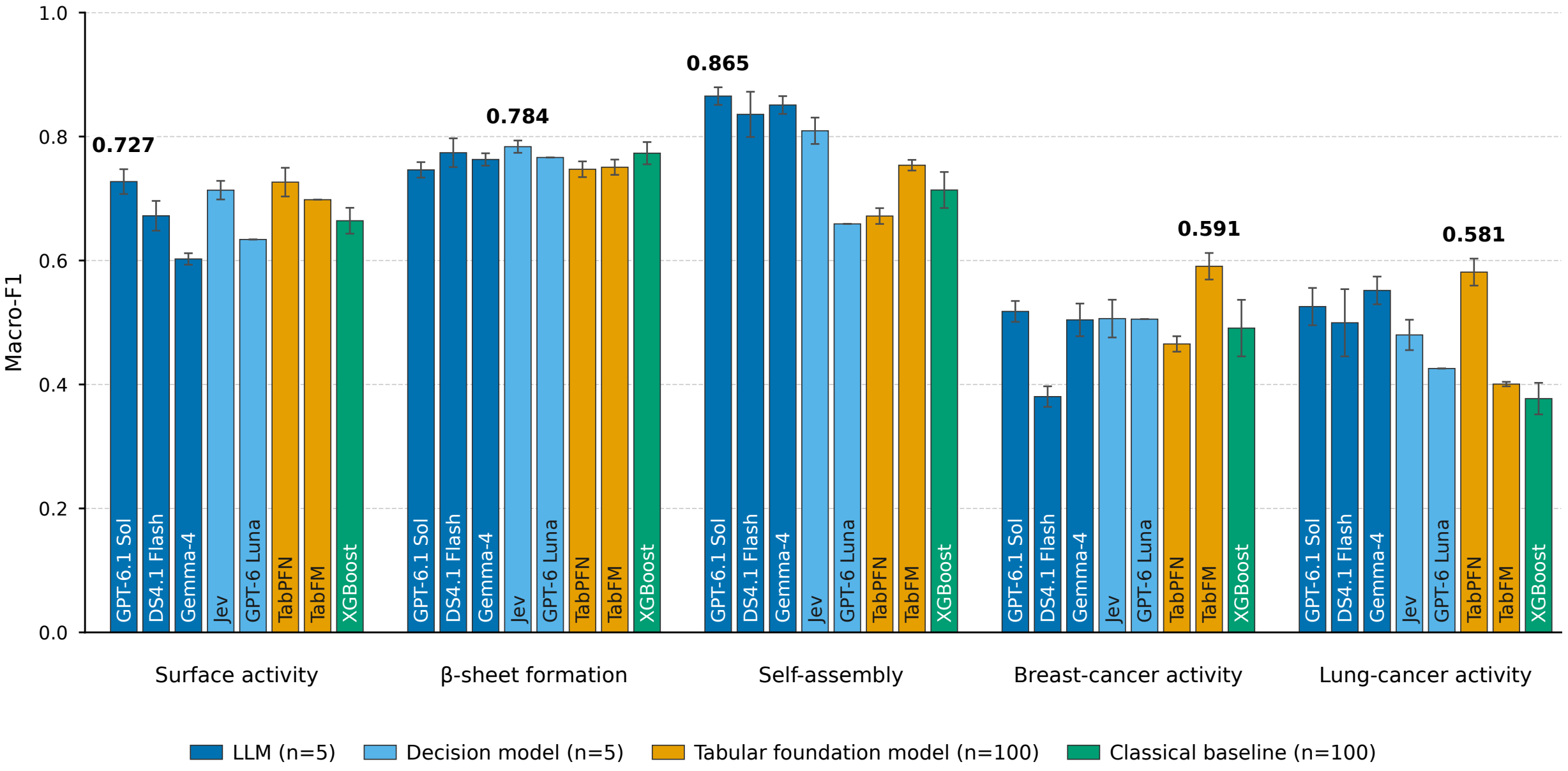

**Figure S1.** Balanced-training performance. Macro-F1 (mean $\pm$ SD) for the eight models on the five tasks after class-balanced training. Each task group contains eight bars, one per model. Bar colour denotes the model family. Error bars show the standard deviation (n = 5 independent repeats for endpoint models; n = 100 seeded repeats for local models). The bold number above the best bar in each group gives its macro-F1.

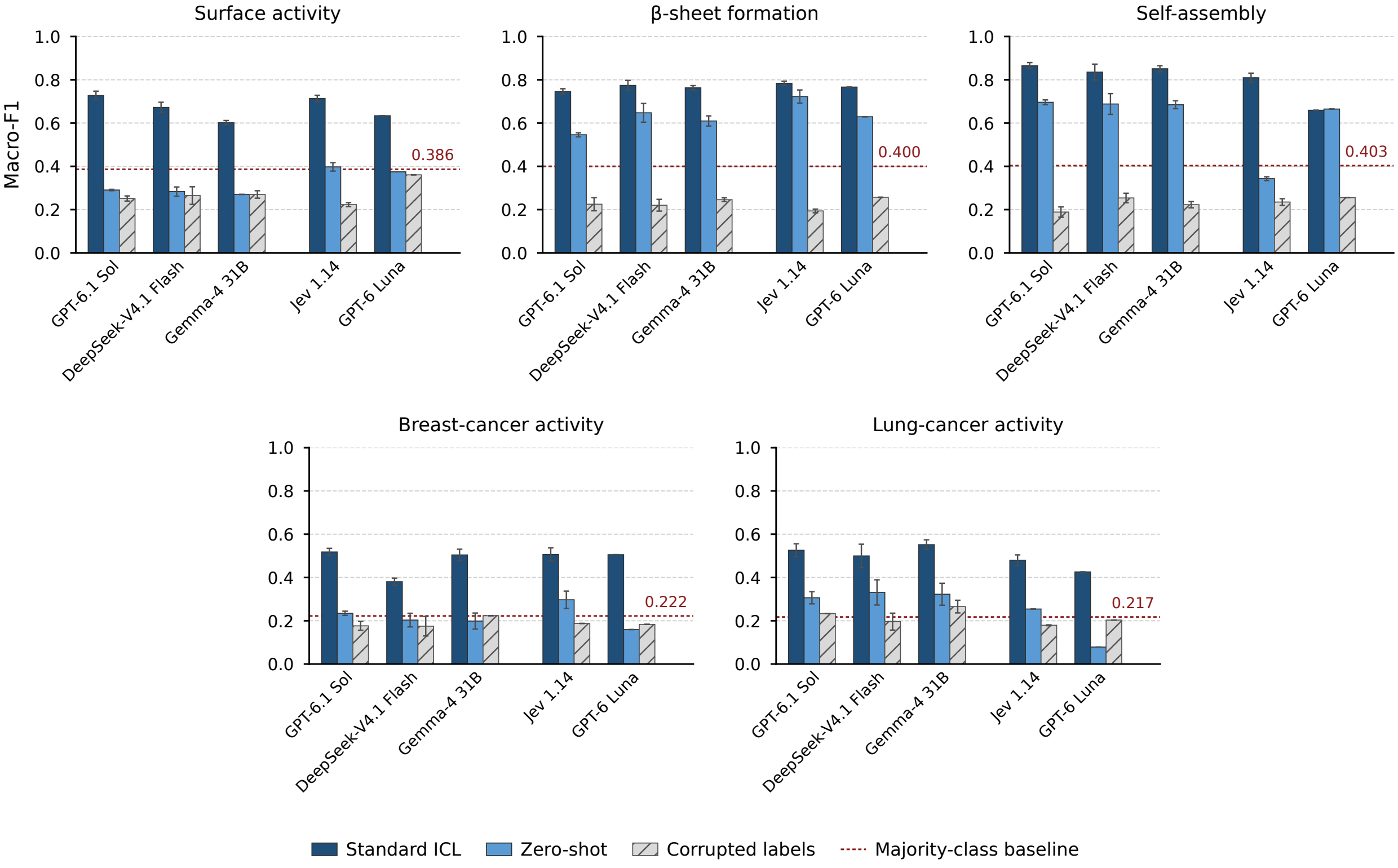

**Figure S2.** Balanced label controls. Macro-F1 for the five endpoint models on the five tasks with standard in-context learning (dark blue), zero-shot prediction (light blue), and corrupted class-balanced labels (hatched). Panels correspond to tasks, and bars are grouped by model. Dashed horizontal lines give the macro-F1 of a majority-class predictor. This predictor always selects the most frequent test label. Standard in-context learning uses the class-balanced reference set. The vertical axis is raw macro-F1, and error bars show the standard deviation across n = 5 independent repeats.

## Bibliography

(1) Houx, J.; Kong, M. J. W.; Tao, S.; Dankloff, P. F. J.; Martens, G.; Hoeksel, J.; Robinson, W. E.; Huck, W. T. S.; Mabesoone, M. F. J. Data-Driven Discovery of pH-Dependent Surfactant Behavior in a Peptide Amphiphile Library, 2026. https://doi.org/10.26434/chemrxiv.15010226.v1.

(2) Talluri, Y. N.; Sankaranarayanan, S. K.; Fry, H. C.; Batra, R. Discovery of Unconventional and Nonintuitive Self-Assembling Peptide Materials Using Experiment-Driven Machine Learning. *Science Advances* **2025**, *11* (24), eadt9466. https://doi.org/10.1126/sciadv.adt9466.

(3) Njirjak, M.; Žužić, L.; Babić, M.; Janković, P.; Otović, E.; Kalafatovic, D.; Mauša, G. Reshaping the Discovery of Self-Assembling Peptides with Generative AI Guided by Hybrid Deep Learning. *Nature Machine Intelligence* **2024**, *6* (12), 1487–1500. https://doi.org/10.1038/s42256-024-00928-1.

(4) Grisoni, F.; Neuhaus, C. S.; Gabernet, G.; Müller, A. T.; Hiss, J. A.; Schneider, G. Designing Anticancer Peptides by Constructive Machine Learning. *ChemMedChem* **2018**, *13* (13), 1300–1302. https://doi.org/10.1002/cmdc.201800204.